\documentclass[12pt]{iopart}
\usepackage{epsf}   \usepackage{epsfig} \usepackage{color}

\definecolor{important}{rgb}{1,0,0}
\newcommand{\bra}[1]{\mbox{$\langle #1 |$}}
\newcommand{\ket}[1]{\mbox{$| #1 \rangle$}}

\begin{document}
\title{Entanglement dynamics in chains of qubits with noise and disorder}
\author{D.I. Tsomokos$^1$, M.J. Hartmann$^{2,3}$, S.F. Huelga$^1$ and M.B. Plenio$^{2,3}$}
\address{$^1$School of Physics, Astronomy \& Mathematics,
University of Hertfordshire, Hatfield, Herts, AL10 9AB, UK}
\address{$^2$ QOLS, Blackett Laboratory, Imperial College London,
Prince Consort Road, London SW7 2BW, UK}
\address{$^3$ Institute for Mathematical Sciences, Imperial College London,
53 Exhibition Road, London SW7 2PG, UK}
\ead{d.i.tsomokos@physics.org (corresponding author),
m.hartmann@imperial.ac.uk}

\date{\today}

\begin{abstract}
The entanglement dynamics of arrays of qubits is analysed in the
presence of some general sources of noise and disorder. In
particular, we consider linear chains of Josephson qubits in
experimentally realistic conditions. Electromagnetic and other
(spin or boson) fluctuations due to the background circuitry and
surrounding substrate, finite temperature in the external
environment, and disorder in the initial preparation and the
control parameters are embedded into our model. We show that the
amount of disorder that is typically present in current
experiments does not affect the entanglement dynamics
significantly, while the presence of noise can have a drastic
influence on the generation and propagation of entanglement. We
examine under which circumstances the system exhibits steady state
entanglement for both short ($N<10$) and long ($N>30$) chains and
show that, remarkably, there are parameter regimes where the
steady state entanglement is strictly non-monotonic as a function
of the noise strength. We also present optimized schemes for
entanglement verification and quantification based on simple
correlation measurements that are experimentally more economic
than state tomography.
\end{abstract}

\pacs{74.50.+r, 03.67.Hk, 05.50.+q} \maketitle

\section{Introduction}

A fundamental property of the superconducting state is that it
exhibits quantum coherence at the macroscopic scale, a feature
that has been used to probe the validity range of quantum
mechanics beyond the microscopic realm \cite{Leggett, newleggett}.
The development of quantum information science and the
experimental progress in the manufacturing and control of
superconducting-based quantum circuits has allowed for novel
proposals aimed at implementing quantum information processing
using {\em Josephson qubits} \cite{reviews}. This generic
denomination refers to qubit realizations that involve the
charge~\cite{Nakamura} or the flux~\cite{Mooij} degree of freedom
in superconducting devices (also see
References~\cite{Nori,Vion,Yu,Wallquist}). The coherent coupling
of two charge qubits and the implementation of conditional gate
operations~\cite{Pashkin}, as well as the coupling of two flux
qubits~\cite{Majer}, have been demonstrated experimentally, and
there is currently an increasing activity in the field.
Interesting applications include proposals to interface such
devices with optical elements in order to create hybrid
technologies~\cite{hybrid}; or to use them for quantum
communication~\cite{Romito,Bose 03,Plenio HE 04}. It needs to be
realised however that the technological barriers for full scale
quantum computation are formidable. Thus there is a need for
intermediate experiments that are interesting and non-trivial yet
less demanding than implementing quantum computation. The
exploration of many-body dynamics provides such a platform.
Crucially, the fabrication of arrays that involve $N \sim 50$
Josephson qubits has already been achieved in the
laboratory~\cite{discussion_Mooij}, and one of our aims in the
present work is to make realistic predictions about their
dynamical entanglement properties. In order to do so, we will take
into account the influence of certain general forms of noise and
disorder on the system.

A well-understood source of noise in any Josephson device is due
to electromagnetic fluctuations in its background circuitry
\cite{noise_circuit}. In the case of a single superconducting
qubit, generic spin or boson fluctuations with a variety of
spectral properties can be treated within the framework of the
spin-boson model \cite{noise_spin-boson}. Note, however, that the
precise sources of $1/f$-type noise have yet to be identified and
that the spin-boson formalism ceases to be valid in the limit of
strong coupling to environmental fluctuators \cite{noise_1/f}.
Moreover, the influence of noise on $N>2$ coupled Josephson qubits
remains largely unexplored \cite{noise_N}. In relation to quantum
information processing, it is important to characterise the
necessary conditions for preserving coherence in a noisy
environment before further steps can be taken in the direction of
designing error correction schemes and (subsequently) fault
tolerant superconducting architectures.

In this paper we formulate an initial model for Josephson-qubit
chains in realistic environments, taking into account the most
common sources of noise. First we analyse the quantum dynamics in
ideal conditions and then discuss the modifications one should
expect when (i) disorder is taken into account and (ii) the system
couples linearly to an environment that is modelled as a bath of
harmonic oscillators, as detailed in section 4. To corroborate our
view that the findings for shorter chains are generic, we also
perform simulations for longer chains with $N\sim 50$ qubits. The
simulations are performed using a time-dependent Density Matrix
Renormalization Group (DMRG) technique \cite{DMRG}, employing a
code previously developed and tested in \cite{HRP06}. Given that
our interest focuses on the study of quantum coherence, the system
dynamics is characterized in terms of entanglement creation as
well as entanglement propagation along the chain. There is
currently no unique way to quantify entanglement in mixed states
(see Reference \cite{ent_review} for a recent review). We choose
the logarithmic negativity~\cite{log_negativity,ent_review}
largely for its ease in computation and the availability of an
operational interpretation \cite{APE03}. It is defined as
\begin{equation} \label{log_neg}
E_{\rm N}(\rho_{i,j}) \equiv \log_2 ||\rho_{i,j}^{T_i}|| \, ,
\end{equation}
where $||.||$ denotes the trace norm of a matrix and
$\rho_{i,j}^{T_i}$ is the partial transpose of the reduced density
matrix $\rho_{i,j}$ for two subsystems $i,j$. Other measures would
give broadly equivalent results \cite{ent_review}.

Another fundamental problem concerns the development of techniques
that allow for the detection and quantification of the
entanglement that is present in a network of qubits. Exciting
experimental progress in this direction for the case of Josephson
qubits has been reported very recently \cite{Martinis}, whereby
the entanglement was demonstrated via full state tomography. As
the latter is a costly and time consuming experimental technique,
strategies aimed at establishing a lower bound on entanglement by
means of determining spin-spin correlations have been developed
\cite{AP06}. We test the performance of these concepts in the
present case and find that, using some optimisation, they provide
very accurate estimates for the amount of entanglement present in
the system.

\section{Entanglement Dynamics under Ideal Conditions}
We consider an open chain of $N$ qubits with nearest-neighbour
interactions. The Hamiltonian of the system is
\begin{eqnarray} \label{H_nq}
{\cal H}_{\rm S} = - \frac{1}{2} \sum_{i=1}^{N} \left(\epsilon_i
\sigma_{i}^{z} + \Delta_i \sigma_{i}^{x} \right) - \frac{1}{2}
\sum_{i=1}^{N-1} K_{i} \sigma^{z}_i \otimes \sigma^{z}_{i+1}
\end{eqnarray}
where $\sigma^{x,y,z}_{i}$ denote Pauli matrices for qubit $i$,
and $K_i$ is the strength of the coupling between nearest
neighbours $i,i+1$ (we set $\hbar = k_B = 1$ throughout). The
control parameters are the energy bias $\epsilon_i$ and the
tunnelling splitting $\Delta_i$. We consider, as an example,
charge qubits~\cite{reviews}, in which case we have $\epsilon_i =
4 E_C(1-2N_g)$ and $\Delta_i = E_J$, where $E_C$ is the charging
energy, $E_J$ is the Josephson energy, and $2eN_g= C_g V_g$ is the
gate charge, which is controlled by the gate capacitance $C_g$ and
voltage $V_g$. Charge qubits are operated in the regime where $E_C
\gg E_J$; therefore the energy scale is set by $E_C$, which was of
the order of $1~{\rm K}$ in the experiment of Reference
\cite{Nakamura}, and we let $E_J/E_C=0.1$. We consider purely
capacitive coupling between the charge qubits~\cite{Wallquist},
and hence the $\sigma^{z}_i \otimes \sigma^{z}_{i+1}$ interaction
dominates. We assume (this condition will be relaxed later on)
that the effective charge number of each qubit is $N_g=1/2$ (i.e.,
$\epsilon_i=0$) so that it is operated at the so-called degeneracy
point \cite{Vion}. As it will become clear later, this choice can
be advantageous when trying to minimise the impact of noise.

A feasible way to achieve generation of entanglement in coupled
many-body systems is non-adiabatic switching of interactions as
demonstrated in harmonic chains \cite{ent_generation,Plenio HE
04}. This approach has the advantage of only requiring moderate
control over the parameters of individual subsystems. In our study
here we will quantify the amount of entanglement that can be
obtained in this way for the model Equation (\ref{H_nq}). In
particular, we will assume that the interqubit couplings $K_i$ are
initially zero and are then non-adiabatically switched to a finite
value. If one were indeed able to switch off the interqubit
couplings completely, then at absolute zero temperature each qubit
would be prepared in its ground state $\ket{+}$ (when operated at
its optimal point, where $\epsilon_i = 0$). The ground state of
the Hamiltonian of Equation (\ref{H_nq}) for $K_i = 0$ is the
uncorrelated state
\begin{equation} \label{psi}
|\Psi(0)\rangle = \ket{+}^{\otimes N}, \qquad \ket{+} =
\frac{1}{\sqrt{2}} (\ket{\uparrow} + \ket{\downarrow})
\end{equation}
where $\ket{\uparrow},\ket{\downarrow}$ denote the eigenstates of
$\sigma^z$ corresponding to zero or one extra Cooper pair in the
superconducting box. We will thus study the \emph{generation} of
entanglement by evolving the initial state of Equation (\ref{psi})
according to the Hamiltonian (\ref{H_nq}) with $K_i \ne 0$. We
will also study the \emph{propagation} of entanglement
\cite{Romito,Bose 03,HRP06,ent_generation,Plenio HE 04} by
assuming that our initial state is
\begin{equation} \label{phi}
|\Phi(0)\rangle = \ket{\beta}_{12} \otimes \ket{+}^{\otimes N-2},
\qquad \ket{\beta}_{12} = \frac{1}{\sqrt{2}} (\ket{\uparrow
\downarrow} + \ket{\downarrow \uparrow}).
\end{equation}
In this case, the interactions $K_i$ are initially zero but a Bell
state $\ket{\beta}_{12}$ has been created for the first two
qubits. Again, the interactions are instantaneously switched on to
a finite value and the evolution according to the Hamiltonian
(\ref{H_nq}) with $K_i \ne 0$ is studied. The Bell state
$\ket{\beta}_{12}$, shared between the first two qubits in the
chain, is maximally entangled. We note that there is also
entanglement generation while the quantum correlations of
$\ket{\beta}_{12}$ propagate along the chain. We discuss below the
effect of deviations in these initial conditions, due to
non-vanishing initial interactions or static disorder.

\begin{figure} \centering 
\includegraphics[width=14cm]{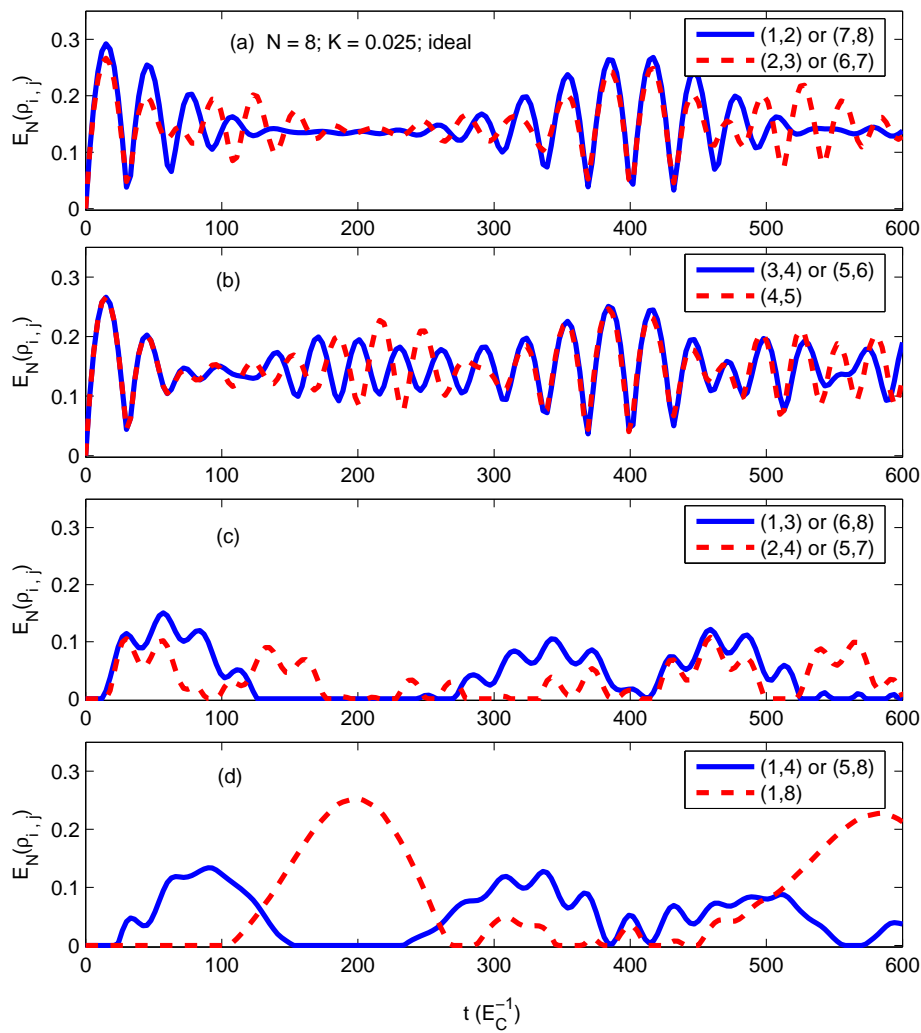}
\caption{Pairwise entanglement creation for a homogeneous chain of
$N=8$ qubits in ideal conditions; the system is described by the
Hamiltonian of Equation (\ref{H_nq}) and the initial state is
$\ket{\Psi(0)}$ of Equation (\ref{psi}). Note that symmetric qubit
pairs, such as (1,2) and (7,8), generate the same entanglement and
hence are represented by the same line. A `collapse-and-revival'
pattern emerges in the entanglement oscillations, as seen from a
comparison of the different panels. \label{Fig_1}}
\end{figure}

We begin by calculating the time-evolution of the logarithmic
negativity of Equation (\ref{log_neg}) for qubit pairs in ideal
conditions. In Figure \ref{Fig_1} we show the result for a chain
of $N=8$ qubits with the initial state $|\Psi(0)\rangle$ of
Equation (\ref{psi}) and parameters $\epsilon_i=0,K_i = \Delta_i /
4$. Due to the geometry of the setup, symmetric pairs of qubits,
such as $(1,2)$ and $(7,8)$ possess the same amount of
entanglement. It is possible to create long-range entanglement,
even between the first and last qubit in the chain (at $t \sim
200~E_{C}^{-1}$ which corresponds to about $1.5~n{\rm s}$). By
comparing and contrasting the four panels in Figure 1 we can see
that there is a characteristic `collapse-and-revival' behaviour:
when the entanglement of nearest or next-nearest neighbours is
constant or vanishing, the entanglement of distant qubits becomes
maximal (e.g., at $t \sim 200~E_{C}^{-1}$). Due to the finite
length of the chain we find that neither the time when a pair of
qubits becomes entangled for the first time, nor the time when the
first entanglement maximum occurs, are proportional to the
distance between the qubits. The characteristic speed at which the
distance of pairwise entanglement is expected to grow is thus
masked by finite size effects, in our study. Entanglement
propagation for this chain in ideal conditions is considered later
(\emph{cf.} Figure \ref{Fig_5}).

{\em Deviations in initial conditions:} In practice, it is not
quite possible to switch off the interqubit couplings completely.
To take this into account we have considered the case when there
is initially some small coupling between the qubits, $K_{\rm
ini}$, and the initial state of the system is the ground state of
${\cal H}_{\rm S}(K_{\rm ini})$. Then the ground state evolves
according to ${\cal H}_{\rm S}(K_{\rm fin})$, where $K_{\rm fin} =
\Delta/4$. In this case we obtain very similar results with those
presented in Figure \ref{Fig_1} for the ideal case (clearly, for
$K_{\rm ini}\rightarrow 0$ we recover the results of the ideal
case). In particular, for $K_{\rm ini} \le \Delta / 100$ the
relative maxima deviate by less than $5\%$, and there is initially
very little entanglement in the system (e.g., the logarithmic
negativity for the first two qubits in the chain at $t=0$ is less
than $0.004$). By contrast, for $K_{\rm ini} \sim \Delta / 10$ the
relative maxima can deviate by up to $30\%$ and the initial
entanglement in the chain is much more evident (e.g., $E_{\rm
N}(\rho_{1,2})\sim 0.1$ at $t=0$ for the same parameters). We will
revisit this case shortly, after we have introduced disorder and
noise into the system ({\em cf.} Figure \ref{Fig_4}).

Another interesting question relating to the initial preparation
concerns the state of the chain at thermal equilibrium. In
particular, we would like to know if we would obtain similar
results when the state of the system at $t=0$ is the thermal
equilibrium state, and also how close are the thermal and ground
states of the system described by the Hamiltonian ${\cal H}_{\rm
S}(K_{\rm ini})$ of Equation (\ref{H_nq}). We have therefore assumed that the
initial state of the chain is the thermal equilibrium state
$\rho(T) = \exp(-\beta {\cal H}_{\rm S})/Z$, where $\beta=1/T$,
$Z={\rm Tr}[\exp(-\beta {\cal H}_{\rm S})]$, and there is initial
coupling, $K_{\rm ini}$, between the qubits. The coupling is then
switched on to its final value $K_{\rm fin}$ at $t=0$. In this
case we have found that for low temperatures ($T \le 20~m{\rm K}$)
the entanglement dynamics of the chain is very similar with that
obtained by evolving the ground state (the relative deviations are
less than $10\%$) for the same value of $K_{\rm ini}$. In order to
compare the thermal equilibrium state $\rho(T)$ with the ground
state $\ket{G}$ of ${\cal H}_{\rm S}(K_{\rm ini})$ we have
calculated the fidelity $\bra{G} \rho(T) \ket{G}$ for various
values of the temperature (with fixed $K_{\rm ini}=\Delta/4$). We
have found that for temperatures $T \le 15~m{\rm K}$ the fidelity
is between $0.99$ and $1$, and hence the two states are very close
indeed for these temperatures. Between $15~m{\rm K}$ and $25~m{\rm
K}$ the thermal equilibrium state and the ground state begin to
differ (their fidelity slowly drops to 0.9 as the temperature is
increased).

%
%

\section{Influence of Disorder}

\begin{figure}[ht] \centering 
\includegraphics[width=\textwidth]{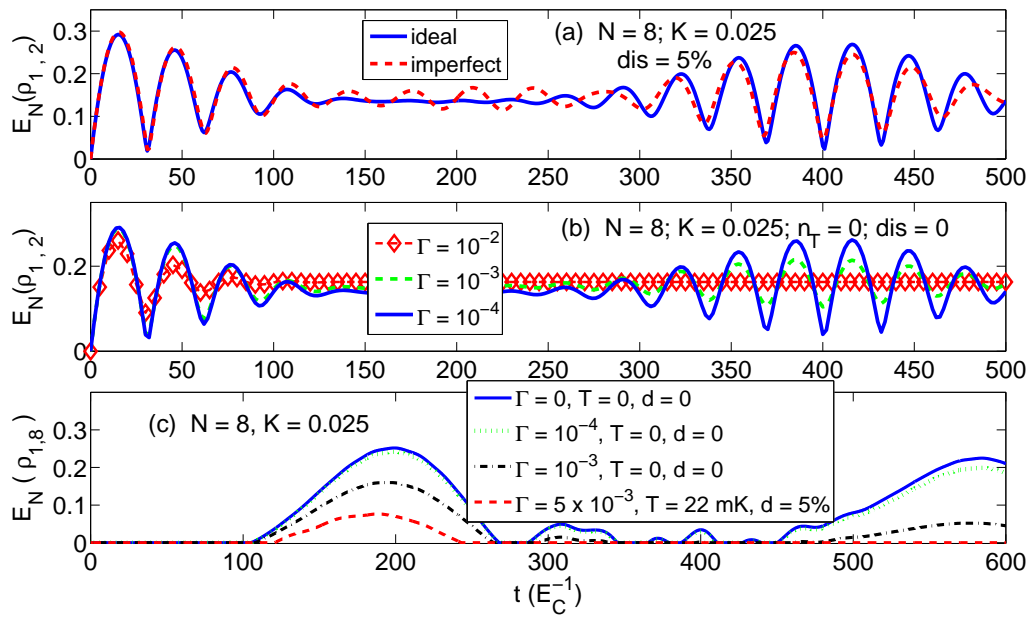}
\caption{Entanglement creation between qubits $(1,2)$ in the
presence of (a) disorder and (b) noise, which is characterised by
a decay rate $\Gamma$ at absolute zero temperature. Subplot (c)
shows $E_{N}(t)$ for qubits (1,8) for various values of the decay
rate $\Gamma$, temperature, and disorder. In (a) the initial state
is $\ket{\Psi(0)}$ of Equation (\ref{psi}) and it evolves under
the Hamiltonian (\ref{H_nq}); in (b) and (c) the initial state is
$\ket{\psi(0)}$ of Equation (\ref{psi_phi}) and it evolves under
the master equation (\ref{master_eq}). \label{Fig_2}}
\end{figure}

In any experimental situation the initial preparation will also
suffer from errors in the control parameters $\alpha_{\rm ctrl} =
\epsilon_i, \Delta_i, K_i$. As a result, the homogeneity of the
chain will be broken. We can simulate the effect by letting the
parameters take random, but static, values in the interval $[(1 -
d)\alpha_{\rm ctrl}, (1 + d)\alpha_{\rm ctrl}]$, where $d$
quantifies the disorder. An example is shown in Figure
\ref{Fig_2}(a), where we plot $E_{\rm N}(t)$ for the pair $(1,2)$
in the ideal (solid line) and imperfect scenario (broken line),
where the disorder in $\Delta_i$ and $K_i$ is $d=0.05$. Averaging
over $10^4$ runs, we have found that disorder with $d = 0.01$, $d
= 0.05$, and $d = 0.10$ causes relative fluctuations of the
maximal entanglement equal to $0.011$, $0.027$ and $0.054$,
respectively. Therefore for disorder which is less than $10\%$
(the upper bound in state-of-the-art experiments
\cite{discussion_Meeson}) the entanglement in the system changes
marginally, on average. This is indeed true for the noisy scenario
also, as shown in the following section. It is noted that disorder
has recently been studied in related, but different, contexts in
\cite{ent_generation,disorder}.

%
%

\section{Noise Model For a Variety of Sources}
We consider a spin-boson Hamiltonian of the form
\begin{equation}
{\cal H}_{\rm SB} = {\cal H}_{\rm S} + {\cal H}_{\rm B} +
\sum_{i=1}^{N} \sigma^{z}_i X_i
\end{equation}
where the first term corresponds to the free system Hamiltonian of
Equation (\ref{H_nq}), the second term is the Hamiltonian for all
independent baths $i=\{1,2,\ldots,N\}$, ${\cal H}_{\rm B} = \sum
_{i=1}^{N} \sum _{k} \Omega^{(i)}_{k} a^{(i)\dagger}_k a^{(i)}_k$,
where the $k$-th mode of bath $i$ has boson creation and
annihilation operators $a^{(i)\dagger}_k$ and $a^{(i)}_k$,
respectively, and the third term is the interaction between a
qubit and its bath, whose `force' operator is $X_i = \sum_{l}
{\cal G}^{(i)}_l \left[a_{l}^{(i)\dagger} + a^{(i)}_{l} \right]$
\cite{noise_spin-boson}. Clearly, it is assumed that each qubit is
affected by its own bath, i.e., $[a^{(i)}_k, a^{(j)\dagger}_k] =
\delta_{ij}$, a reasonable requirement for charge qubits biased by
independent voltage gates.

In the coherent regime, where $\omega_{i} \equiv (\epsilon_{i}
^{2} + \Delta_{i}^{2})^{1/2}$ is much larger than the thermal
energy, the preferred basis is given by the eigenstates of the
single-qubit Hamiltonian, i.e., $\ket{0} = \cos(\theta_i /2) \ket
{\uparrow} + \sin (\theta_i/2) \ket{\downarrow}$ and $\ket{1} =
-\sin (\theta_i /2) \ket{\uparrow} + \cos(\theta_i /2)
\ket{\downarrow}$, where the mixing angle obeys $\tan\theta_i =
\Delta_i / \epsilon_i$. In this basis, ${\cal H}_{\rm SB}$ becomes
\begin{equation}
{\cal H}'_{\rm SB} = {\cal H}'_{\rm S} + {\cal H}_{\rm B} +
\sum_{i=1}^{N} (\sin\theta_i \sigma^{x}_{i} + \cos \theta_i
\sigma^{z}_{i}) X_i
\end{equation}
where
\begin{eqnarray} \label{H_sys}
{\cal H}'_{\rm S} = - \frac{1}{2}\sum_{i=1}^{N} \omega_{i}
\sigma^{z}_{i} - \frac{1}{2}\sum_{i=1}^{N-1} K_{i} (c_i
\sigma^{z}_i + s_i \sigma^{x}_i) (c_{i+1} \sigma^{z}_{i+1} +
s_{i+1} \sigma^{x}_{i+1})
\end{eqnarray}
is the system Hamiltonian (the Pauli matrices are now written in
the $\{\ket{0},\ket{1}\}$ basis) with $c_i = \cos\theta_i,~s_i =
\sin\theta_i$. In this basis the states $\ket{\Psi(0)}$ and
$\ket{\Phi(0)}$ of Equations (\ref{psi}) and (\ref{phi}),
respectively, become
\begin{equation} \label{psi_phi}
|\psi(0)\rangle = \ket{0}^{\otimes N}, \qquad
|\phi(0)\rangle = \ket{\beta'}_{12} \otimes \ket{0}^{\otimes N-2}
\end{equation}
where $\ket{\beta'}_{12} = 2^{-1/2} (\ket{0 1} + \ket{1 0})$.

When the bath's degrees of freedom are traced out, and within the
Born-Markov approximation, the time evolution of the chain is
governed by a master equation of the Lindblad form
\begin{equation}\label{master_eq}
\dot{\rho}=-i[H'_{\rm S},\rho]+{\cal L}\rho
\end{equation}
where $H'_{\rm S}$ is given by the same expression of Equation (7)
provided that the weak coupling limit, where $K_i < \omega_i$,
holds. The damping terms are given by the usual expressions,
\begin{eqnarray}
\fl{ {\cal L}\rho = \sum_{i=1}^{N} [G_i(2 \sigma_{i}^{+} \rho
\sigma_{i}^{-} - \rho \sigma_{i}^{-} \sigma_{i}^{+} -
\sigma_{i}^{-} \sigma_{i}^{+} \rho) + \tilde{G}_i (2
\sigma_{i}^{-} \rho \sigma_{i}^{+}- \rho \sigma_{i}^{+}
\sigma_{i}^{-} - \sigma_{i}^{+} \sigma_{i}^{-} \rho)} \nonumber \\
+ g_i (2 \sigma_{i}^{z} \rho \sigma_{i}^{z} - 2\rho) ]
\end{eqnarray}
where $\sigma_{j}^{\pm} \equiv 2^{-1}(\sigma_{j}^{x} \pm i
\sigma_{j}^{y})$ and the parameters are defined as
\begin{eqnarray}
G_i = \sin^{2} \theta_i (1+n_{\rm T})\Gamma, \; \; \; \;
\tilde{G}_i = \sin^{2} \theta_i n_{\rm T}\Gamma,  \; \; \; \; g_i
= \cos^{2} \theta_i \Gamma
\end{eqnarray}
with $n_{\rm T}$ denoting the average number of bosons in the
environment. We assume that the environments of all qubits are
identical. We have not specified the environment's spectral
properties and hence the {\em decay rate} $\Gamma$ is given as a
phenomenological parameter whose exact value can be adjusted to
match that obtained for the actual spectral density of the bath.
Within this framework, a broad class of dissipative effects can be
accounted for, ranging from electromagnetic fluctuations in the
surrounding circuitry to, for instance, the coupling to a phonon
bath \cite{phonons}. The system can therefore be viewed as a non
critical dissipative Ising chain (The critical behaviour of
dissipative Ising chains has been recently addressed in
\cite{werner}.)

At the degeneracy point $\epsilon_i=0$ and $\cos \theta_i = 0$. As
a result, each qubit is susceptible to relaxation only (the
`optimal' point introduced in Reference \cite{Vion}). If, however,
the energy bias is not exactly zero then the longitudinal
contribution $\sigma^{z}_{i} X_i$ leads to pure dephasing at a
rate $g_i$. In any experimental realization, the presence of
disorder limits the accuracy with which qubits can be operated at
their optimal points, especially when it comes to the operation of
long chains. In what follows we take this into account and study
the modifications due to disorder. In current experiments
\cite{discussion_Meeson} the value of disorder is typically
$5-10\%$ at temperatures $20-40~m{\rm K}$. The decoherence time
$t_{\rm d} \equiv \Gamma^{-1}$ for two coupled charged qubits
\cite{Pashkin} was reported to be around $2.5~n{\rm s}$ (for
single qubits $t_{\rm d}$ can be higher). In our simulations below
we usually assume a worst-case scenario and let $d=5\%$ at $T
\approx 41~m{\rm K}$ with decay rate $\Gamma = 10^{-2} E_{C}$
(which corresponds to $t_{\rm d} \approx 1~n{\rm s}$).

\section{Dynamics of Short Chains $(N \sim 10$ qubits)}

\begin{figure} \centering 
\includegraphics[width=13cm]{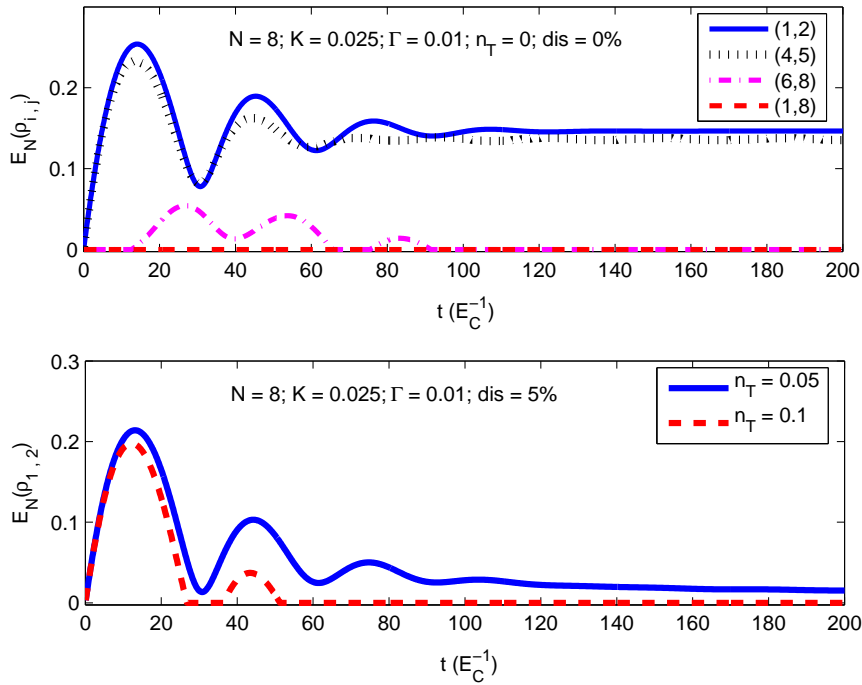}
\caption{Entanglement creation between two qubits for
the case of relaxation only (top) and relaxation, finite
temperature, and disorder (bottom). The initial state is
$\ket{\psi(0)}$ of Equation (\ref{psi_phi}) and the system evolves
under the master equation (\ref{master_eq}). \label{Fig_3}}
\end{figure}

\begin{figure} \centering 
\includegraphics[width=\textwidth]{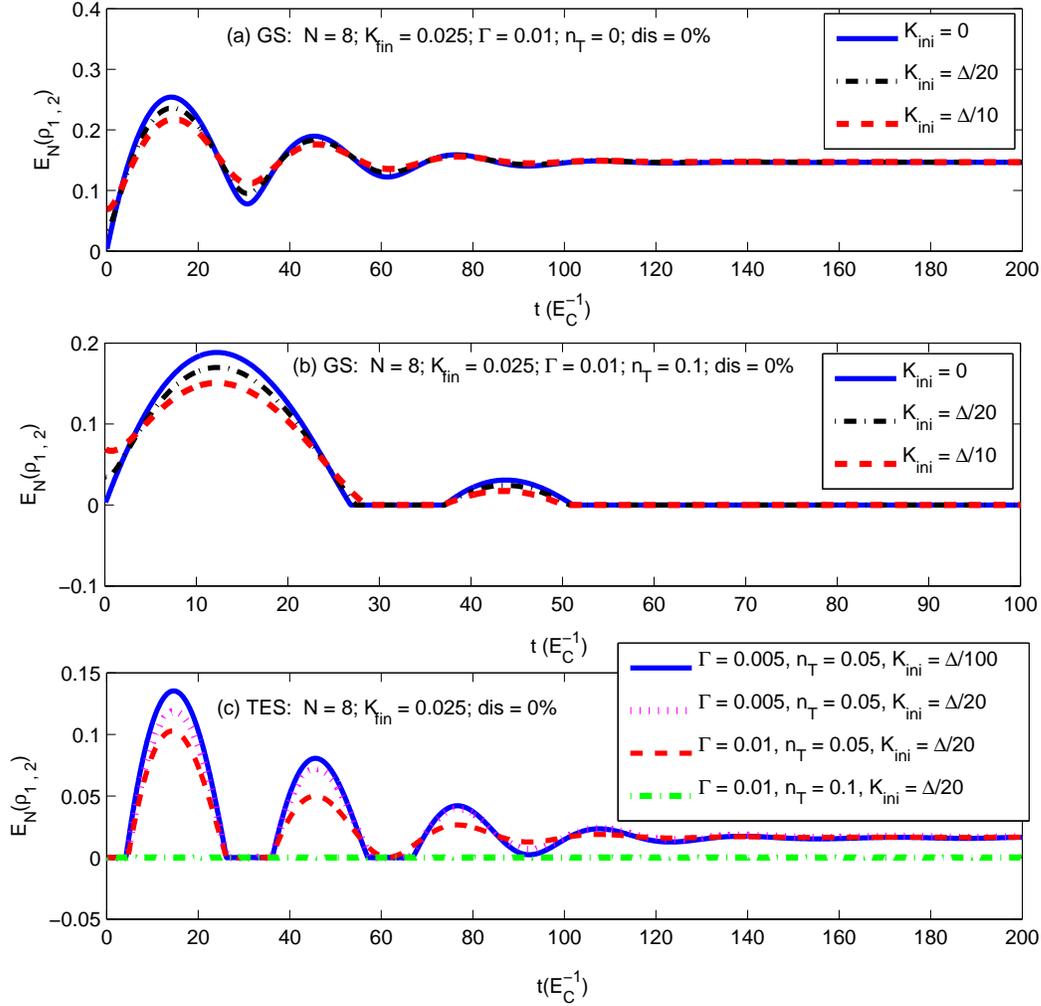}
\caption{Entanglement creation between qubits (1,2) in noisy
conditions when there is some initial homogeneous coupling $K_{\rm
ini}$ that is instantaneously switched on to $K_{\rm
fin}=\Delta/4$ at $t=0$. In subplots (a) and (b) the initial state
is the ground state (`GS') of ${\cal H}'_{\rm S}(K_{\rm ini})$ of
Equation (\ref{H_sys}) and they correspond to temperatures $T=0$
and $T \approx 41~m{\rm K}$ ($n_{\rm T}=0.1$). In subplot (c) the
initial state is the thermal equilibrium state (`TES') of the
system at a given temperature $T$, for various values of the
parameters. \label{Fig_4}}
\end{figure}

\begin{figure} \centering
\scalebox{0.65}{\includegraphics{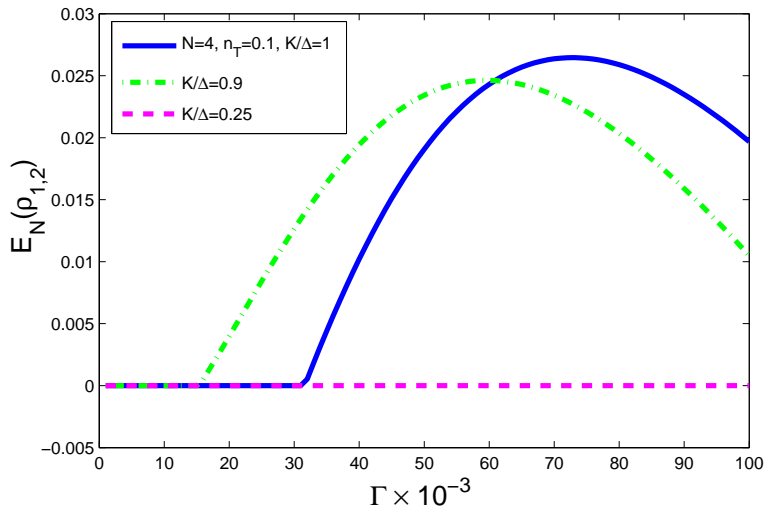}} \caption{Steady state
entanglement between qubits (1,2) in a chain of $N=4$ when $n_{\rm
T}=0.1$ and for different values of the ratio $K/\Delta$ as a
function of the noise strength $\Gamma$. Parameter regimes can be
identified where entanglement generation is enhanced by amplifying
the environmental noise. \label{Fig_5}}
\end{figure}

\begin{figure} \centering 
\includegraphics[width=13cm]{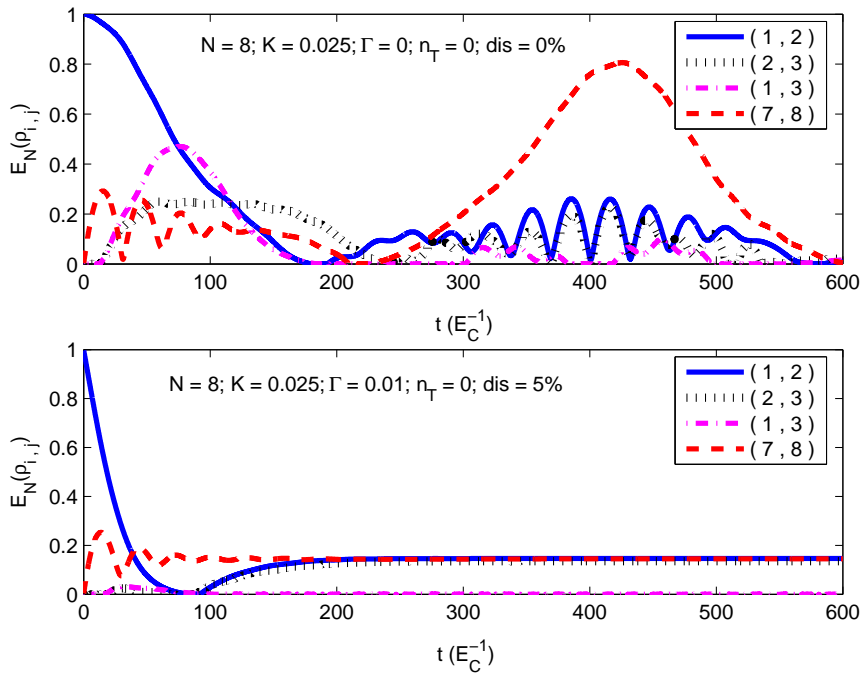}
\caption{Entanglement propagation at zero temperature for the
ideal case (top) and for relaxation and disorder (bottom). In the
top plot, the initial state is $\ket{\Psi(0)}$ of Equation
(\ref{phi}) and the system is described by the Hamiltonian ${\cal
H}_{\rm S}$ of Equation (\ref{H_nq}). In the bottom plot the
initial state is $\ket{\phi(0)}$ of Equation (\ref{psi_phi}) and
it evolves under the master equation (\ref{master_eq}).
\label{Fig_6}}
\end{figure}

In this section we discuss the results on the entanglement
dynamics of open chains with $N=8$ qubits (Figures
\ref{Fig_1}-\ref{Fig_6}). Figures \ref{Fig_1} and \ref{Fig_2}(a)
have been analysed previously. Figure \ref{Fig_2}(b) shows the
creation of entanglement in the pair $(1,2)$ for various values of
the relaxation rate $\Gamma$. Figure \ref{Fig_2}(c) shows the
creation of entanglement in the pair $(1,8)$ for various values of
$\Gamma$ and other parameters. It is seen that for noise strength
$\Gamma=5 \times 10^{-3}$ (i.e., $t_{\rm d} \approx 5~n{\rm s}$),
average number of photons $n_{\rm T} = 0.01$ (i.e., $T\approx
22~m{\rm K}$) and disorder $d=5\%$ one may still obtain
substantial entanglement between the first and last qubit in the
chain (in particular, the ratio of the values of the first maxima
corresponding to the imperfect / ideal cases is approximately
$2/5$).

In Figure \ref{Fig_3} we plot $E_N(t)$ for different pairs of
qubits in the case of (top) relaxation with $\Gamma = 10^{-2}$ at
zero temperature ($T=0$) and (bottom) relaxation with $\Gamma =
10^{-2}$, finite temperature $T \approx 41~m{\rm K}$ ($n_{\rm T} =
0.1$) or $T \approx 33~m{\rm K}$ ($n_{\rm T} = 0.05$), and
disorder $5\%$ ($d = 0.05$). As expected, the entanglement beyond
nearest neighbours is drastically reduced in the presence of
larger values of the noise (the correlations between the first and
last qubit in the chain vanish altogether for this high value of
$\Gamma$).

%
%

In Figure \ref{Fig_4} we study the creation of entanglement
between the first two qubits in the chain (in subplots (a) and
(b)) when the initial state is the ground state of the Hamiltonian
${\cal H}'_{\rm S}(K_{\rm ini})$ of Equation (\ref{H_sys}), for
various values of the initial coupling strength $K_{\rm ini}$. At
$t=0$ the coupling is instantaneously switched on to its final
value $K_{\rm fin} = \Delta / 4$. Subplot (a) shows the case with
noise, at absolute zero temperature. Subplot (b) takes into
account the temperature in the environment ($T\approx 41~m{\rm
K}$). In subplot (c) we study the case whereby at $t=0$ the state
of the system is in thermal equilibrium with its environment at
temperature $T_0$. Therefore in this case we let $\rho(T_0) =
\exp(-\beta{\cal H}'_{\rm S})/Z$ at $t=0$. The system Hamiltonian
${\cal H}'_{\rm S}$, given by Equation (\ref{H_sys}), depends on
the initial interqubit coupling $K_{\rm ini}$. The evolution
proceeds according to the master equation (\ref{master_eq}) with
the coupling $K_{\rm fin}$ and an average number of photons
$n_{\rm T}$ that corresponds to the temperature $T_0$. It is seen
that at operating temperatures of around $40~m{\rm K}$ the
entanglement vanishes. It is however possible to observe
entanglement when the temperature gets smaller (e.g., for
$T_0\approx 33~m{\rm K}$).

%
%
%

The results in Figure \ref{Fig_4} seem to indicate that the
increase in the noise strength $\Gamma$ and the external
temperature yield the unavoidable degrading of entanglement
generation. The amplitude of the entanglement oscillations
decreases and the system becomes separable in the steady state for
$\Gamma$ and/or $n_{\rm T}$ sufficiently large. However, this
behaviour is not universal and we need to differentiate two time
scales in the system.  The initial transient is always such that
the amplitude of entanglement oscillations is reduced as the noise
increases and the amplitude of the first entanglement maximum is a
monotonically decreasing function of both $\Gamma$ and $T$.
However, for a fixed $n_{\rm T}$, the steady-state entanglement
can display a non-monotonic behaviour as a function of $\Gamma$.
This phenomenon is of the same type of the noise-assisted effects
that have been studied in Reference \cite{hp} for weakly driven
spin chains and is illustrated in Figure \ref{Fig_5} for a system
of $N=4$ qubits. We see that at the selected temperature where
$n_{\rm T}=0.1$, there are parameter regimes for which the
steady-state entanglement is initially zero for low values of the
noise strength and resurfaces when $\Gamma$ is increased over a
certain threshold. This result indicates that if the aim is to
generate entanglement in the steady state, it may be advantageous
to amplify the environmental noise so as to maximise entanglement
production along the chain. Persistence of this effect in longer
chains $(N \sim 40$) has been corroborated numerically.

%
%

Propagation of entanglement is analysed in Figure \ref{Fig_6} for
(a) ideal and (b) non-ideal conditions. In the ideal case,
entanglement propagates from the first two qubits to the last two,
but not perfectly. When we take into account noise and disorder
the entanglement transfer is not possible and the last two qubits
quickly reach their steady state, which is slightly entangled at
absolute zero temperature, but separable at $T \sim 20 ~m{\rm K}$
for the selected parameter regime.

%
%

\section{Dynamics of Long Chains}

To confirm the validity of our findings for longer chains, we have
performed time-dependent DMRG simulations \cite{DMRG}.

For an ideal chain without noise and disorder, we have considered
entanglement generation in the model (\ref{H_nq}) with $N=20$
qubits. Here the matrix dimension was chosen $dim=20$ and a 4th
order Suzuki-Trotter decomposition was employed. The results are
in good agreement with the findings for shorter chains in
figure~\ref{Fig_1}. In particular, we observe the same
`collapse-and-revival' pattern of the entanglement oscillations,
and the long-range entanglement peaks at those regions where the
short-range entanglement is close to its steady-state value or
vanishes altogether.

For the cases which include noise and disorder, a matrix product
representation for mixed states with matrix dimension $dim=60$ and
a 4th order Suzuki-Trotter decomposition were used for a chain of
$N=40$ qubits \cite{DMRG}. A sketch of the method is given in
\ref{DMRGapp}.

\begin{figure} \centering
\includegraphics[width=\textwidth]{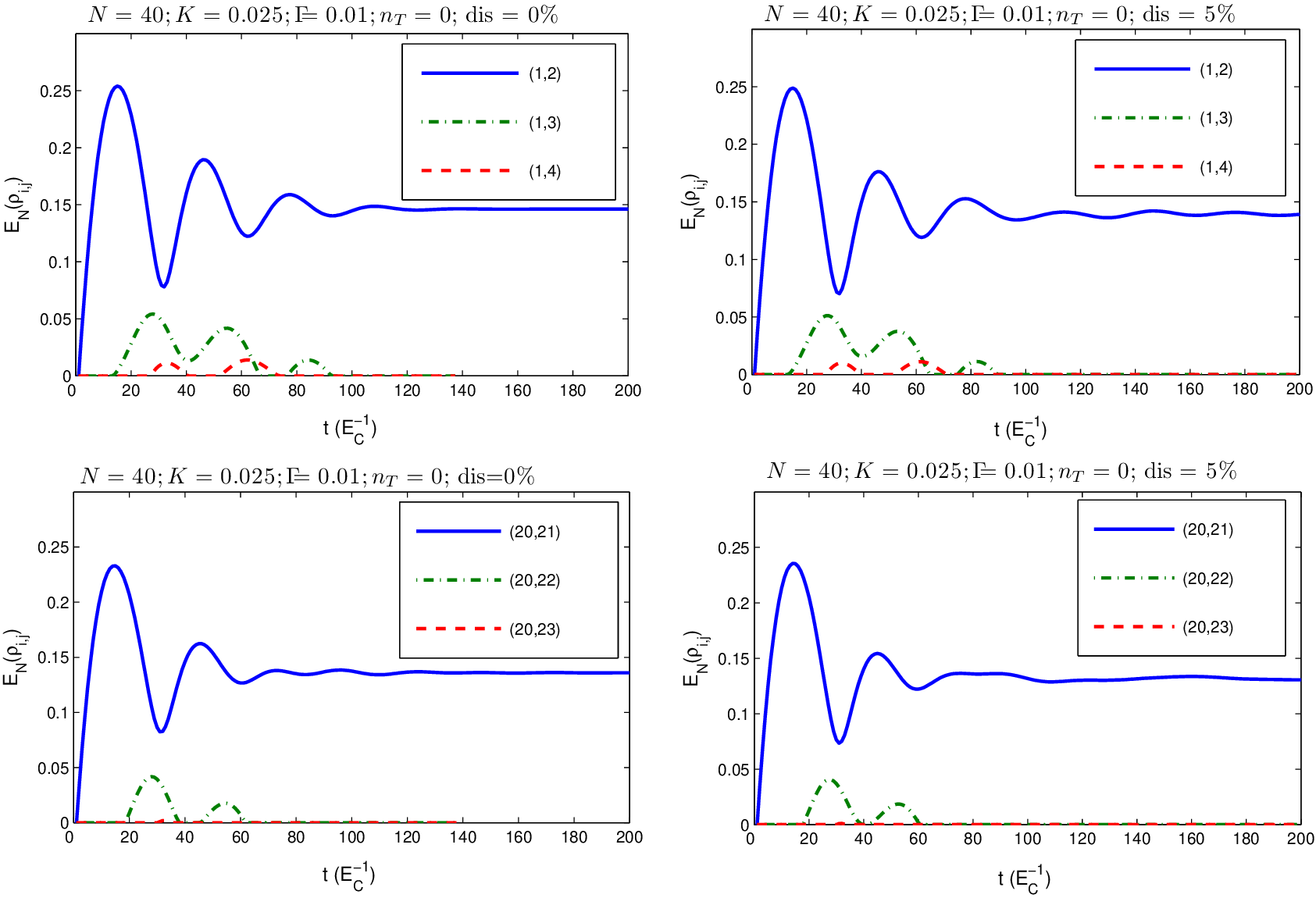}
\caption{Entanglement creation, i.e. evolution of the initial
state $\ket{\psi(0)}$, (\ref{psi_phi}), given by Equation
(\ref{master_eq}) in a chain of $N = 40$ qubits at zero
temperature, in the presence of noise ($\Gamma = 0.01$). The two
plots on the left show the homogeneous case, while the two plots
on the right show a case with $5\%$ disorder in
$\epsilon_i,\Delta_i,K_i$. Qubits at the boundaries are slightly
stronger entangled than in the centre of the chain. The
entanglement between qubits that are further apart than shown here
is zero. \label{homdislongfig}}
\end{figure}

\begin{figure} \centering
\includegraphics[width=12cm]{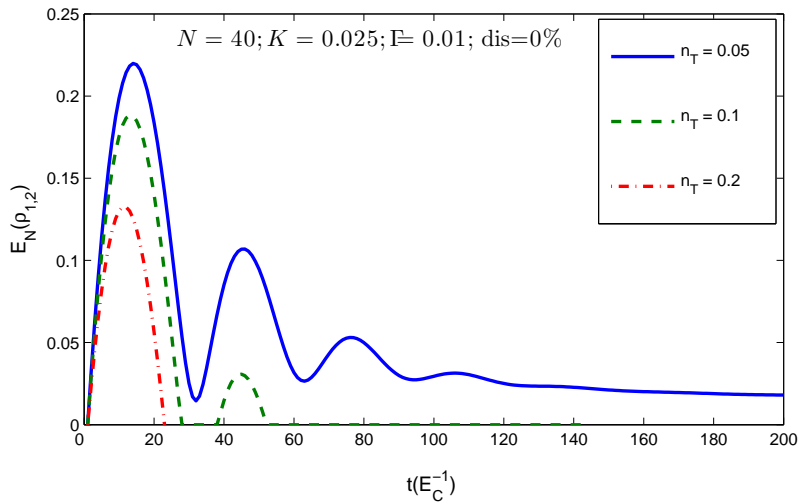}
\caption{Entanglement creation, i.e. evolution of the initial
state $\ket{\psi(0)}$, (\ref{psi_phi}), given by equation
(\ref{master_eq}) in a chain of $N = 40$ qubits at various
temperatures ($n_{\rm T} = 0.05; n_{\rm T}= 0.1$ and $n_{\rm T} =
0.2$) in the presence of noise ($\Gamma = 0.01$). Only nearest
neighbours become entangled in this case.
\label{nonzerotemplongfig}}
\end{figure}

Figure \ref{homdislongfig} shows the creation of entanglement in
the presence of noise, at zero temperature for both a homogeneous
and a disordered chain (in which case disorder occurs in
$\epsilon_i$ as well as $\Delta_i, K_i$). Figure
\ref{nonzerotemplongfig} shows entanglement creation in a noisy
homogeneous chain for various values of temperature. For all
quantities we find good agreement with the results obtained for $N
= 8$, where the relative deviations between $N = 40$ and $N = 8$
are less than $5\%$. It is also noted that the entanglement
between two blocks of two qubits each was found to be about $17\%$
higher than the entanglement between individual qubits of the same
separation.

%
%

%
%

\section{Witnessing Quantum Correlations: Experimental Verification of Entanglement}

In experiments it will be crucial to verify the existence of
entanglement via measurements, which ideally should also permit a
quantification of the detected entanglement. This could be done by
full state tomography, which is a very costly experimental
procedure though. Being able to establish a lower bound on
entanglement from the measurement of a few observables will thus
be a significant advantage. Recently, a theoretical framework for
the exploration of these questions has been developed for general
observables \cite{AP06} and witness operators \cite{EBA06}.

The basic approach is to identify the least entangled quantum
state that is compatible with the measurement data. The
entanglement of that state then provides a quantitative value for
the entanglement that can be guaranteed given the measurement
data. In \cite{AP06}, in particular, spin-spin correlations have
been used to determine such a lower bound analytically. We now
employ this concept for our system and consider the two
quantities,
\begin{eqnarray} \label{bounds}
C_1(\rho_{i,j}) & \equiv & {\rm max}
\left[0, \log_2\left( |C^{xx}_{i,j}| + |C^{zz}_{i,j}| \right) \right] \\
C_2(\rho_{i,j}) & \equiv & {\rm max} \left[0, \log_2\left(1 +
|C^{xx}_{i,j}| + |C^{yy}_{i,j}| + |C^{zz}_{i,j}|\right) - 1\right]
\end{eqnarray}
where $C^{ab}_{i,j} = {\rm Tr}[\sigma_i^a \sigma_j^b \rho]$ ($a,b
= x,y,z$). Both quantities form a lower bound to the logarithmic
negativity, i.e. $E_{\rm N}(\rho_{i,j}) \ge C_1(\rho_{i,j})$ and
$E_{\rm N}(\rho_{i,j}) \ge C_2(\rho_{i,j})$. $C_1(\rho_{i,j})$ can
be employed if only $C^{xx}_{i,j}$ and $C^{zz}_{i,j}$ are
accessible in measurements. However, if $C^{yy}_{i,j}$ can be
measured too, then $C_2(\rho_{i,j})$ yields a tighter bound.

\begin{figure} \centering
\includegraphics[width=12cm]{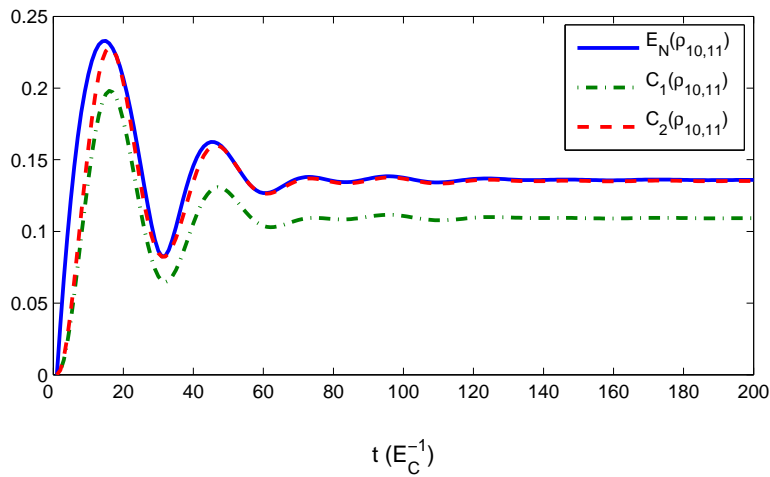}
\caption{The logarithmic negativity (\ref{log_neg}) and the two lower bounds
(\ref{bounds}) for a chain of $N = 40$ qubits with $\Delta = 0.1$,
$K = 0.025$ and $\Gamma = 0.01$ at $n_{\rm T} = 0$.
\label{correlfig}}
\end{figure}

Figure \ref{correlfig} shows that both lower bounds provide good
approximations for the logarithmic negativity of two neighbouring
qubits. If the qubits are next-nearest neighbours,
$C_2(\rho_{i,j})$ still provides a good estimate, while
$C_1(\rho_{i,j})$ eventually fails to approximate the entanglement
well.

%
%

The reason why $C_1(\rho_{i,j})$ and $C_2(\rho_{i,j})$ sometimes
do not approximate the entanglement very well lies in the choices
of the axes along which correlations are measured. Instead of
$C^{xx}_{i,j}$, $C^{yy}_{i,j}$ and $C^{zz}_{i,j}$ one could
consider correlations along a rotated set of axes, $C^{aa}_{i,j}$,
$C^{bb}_{i,j}$ and $C^{cc}_{i,j}$, where $\sigma_i^a =
\sum_{\alpha = x,y,z} R_{a \alpha} \sigma_i^{\alpha}$ and $R_{a
\alpha}$ is an orthogonal matrix representing the rotation.
Choosing to measure correlations along $x$, $y$ and $z$ may hence
underestimate the entanglement severely. The best approximation of
the entanglement is obtained by maximising $C^{aa}_{i,j}$,
$C^{bb}_{i,j}$ and $C^{cc}_{i,j}$ over all possible choices for
the axes $a$, $b$ and $c$.

This optimal choice can be obtained in the following way: If the
state $\rho$ is symmetric with respect to subsystems $i$ and $j$
in the sense that $C^{xy}_{i,j} = C^{yx}_{i,j}$, $C^{xz}_{i,j} =
C^{zx}_{i,j}$ and $C^{yz}_{i,j} = C^{zy}_{i,j}$, then the matrix
\begin{equation}
X = \left(
\begin{array}{ccc}
C^{xx}_{i,j} & C^{xy}_{i,j} & C^{xz}_{i,j}\\
C^{yx}_{i,j} & C^{yy}_{i,j} & C^{yz}_{i,j}\\
C^{zx}_{i,j} & C^{zy}_{i,j} & C^{zz}_{i,j}
\end{array}
\right)
\end{equation}
is real and symmetric and hence has real eigenvalues and is
diagonalised by a rotation. Let us denote the eigenvalues of $X$
by $\lambda_1$, $\lambda_2$ and $\lambda_3$, then the quantity
\begin{equation}
C'_2(\rho_{i,j}) \equiv {\rm max}\left[0, \log_2\left(1 +
|\lambda_1| + |\lambda_2| + |\lambda_3|\right) - 1\right]
\end{equation}
provides the best approximation of $E_{\rm N}(\rho_{i,j})$ of the
form (\ref{bounds}), as $\lambda_1$, $\lambda_2$ and $\lambda_3$
are the spin-spin correlations along the optimal choice of axes
\footnote{According to theorem VIII.3.9 of \cite{Bhatia}, for a
square matrix $X$, $\sum_i |X_{ii}| \le \sum_i |\lambda_i|$, where
the $\lambda_i$ are the eigenvalues of $X$. As $\sum_i
|\lambda_i|$ is a unitarily invariant matrix norm, it does not
depend on the choice of basis while $\sum_i |X_{ii}| $ does. Thus
the largest value that can be achieved for $\sum_i |X_{ii}| $ is
given by $\sum_i |\lambda_i|$.}.

As an example, in figure \ref{correlfig}, the entanglement between
qubits 10 and 11 is $E_{\rm N}(\rho_{10,11}) = 0.2096$ at $t = 10
E_C^{-1}$. While $C_2(\rho_{10,11}) = 0.1583$ at this point, we
obtained $C'_2(\rho_{10,11}) = 0.2096$ for the optimal choice of
axes $a$, $b$ and $c$. The optimal choice of axes depends on time.
Yet one fixed set of axes approximated the entanglement very well
over a range of $\Delta t = 5 E_C^{-1}$ in our example.

\section{Conclusions}

We have studied the dynamics of entanglement in qubit chains
influenced by noise and static disorder. This study provides
useful analysis for interesting experiments with quantum devices
that in the long term may be suitable for the implementation of
quantum computing. We have considered an experimentally
interesting implementation using Josephson charge qubits with
capacitive interactions between nearest neighbours. We have found
that static disorder less than $10\%$ (i.e., the current
experimental upper bound) does not affect the entanglement
dynamics substantially. By contrast, the influence of
environmental noise, modelled here as a set of independent
harmonic oscillator baths of arbitrary spectral density, is much
more pronounced: it reduces long-range correlations and decreases
the magnitude of the achievable bipartite entanglement. For
typical operating temperatures, the influence of noise on the
chain dynamics at short times and in the steady-state can be
crucially distinct; while the entanglement amplitudes in the
initial transient decrease monotonically with the noise strength,
the steady-state response is non-monotonic. Therefore, we have
identified parameter regimes in which the bipartite entanglement
increases as a result of amplifying the noise. We have found
agreement between the behaviour of entanglement in short ($N\sim
10$) and long ($N\sim 50$) chains.

The present results are encouraging from the experimental point of
view as they suggest that (a) both short- and long-range
entanglement can be generated and propagated under suitable
laboratory conditions, and (b) lower bounds can be placed on the
entanglement from the measurement of a few general observables.

\ack
We are grateful to Rosario Fazio, Hans Mooij and Phil Meeson for
stimulating discussions. This work was supported by the EPSRC -
IRC on Quantum Information and EP/D065305/1, the EU via the
Integrated Project QAP (`Qubit Applications') and the Royal
Society. MJH is supported by the Alexander von Humboldt
foundation. MBP holds a Wolfson Research Merit Award.
%

\appendix
\section{Matrix Product State Simulations for Mixed States
\label{DMRGapp}}

Here we outline the concept proposed in \cite{DMRG} and its
adaption to our application. For the Matrix Product State
simulation of mixed state dynamics, the density matrix for $N$
qubits is expanded in a basis of matrices formed by direct
products of the elementary matrices
\begin{eqnarray}
\epsilon_1 = \left( \begin{array}{cc} 1 & 0 \\ 0 & 0 \end{array} \right); \quad
\epsilon_2 = \left( \begin{array}{cc} 0 & 1 \\ 0 & 0 \end{array} \right); \\
\epsilon_3 = \left( \begin{array}{cc} 0 & 0 \\ 1 & 0
\end{array} \right); \quad \epsilon_4 = \left( \begin{array}{cc} 0 & 0 \\ 0 & 1
\end{array} \right).
\end{eqnarray}
Hence the matrices forming the basis for $N$ qubits are of the
form
\begin{equation}
\underbrace{\epsilon_i \otimes \epsilon_j \otimes \dots \otimes
\epsilon_l}_{N~{\rm sites}}.
\end{equation}
The expansion of the density matrix $\rho$ is now written in terms
of products of matrices in the following way:
\begin{equation}
\fl \rho = \sum_{s_1, s_2, \dots s_N = 1}^4 \, \Gamma_1^{[s_1]}
\cdot \Lambda_1 \cdot \Gamma_2^{[s_2]} \cdot \Lambda_2 \cdot \dots
\cdot \Lambda_{N-1} \cdot \Gamma_N^{[s_N]} \: \epsilon_{s_1}
\otimes \epsilon_{s_2} \otimes \dots \otimes \epsilon_{s_N}
\end{equation}
where `$\cdot$' denotes matrix multiplication. Here, each
$\Gamma_1^{[s_1]}$ ($s_1 = 1,2,3,4$) is a row vector of length
$D$, each $\Gamma_N^{[s_N]}$ is a column vector of length $D$,
each $\Gamma_j^{[s_j]}$ ($j \not= 1, N$) is a $D \times D$ matrix
and each $\Lambda_j$ is a diagonal $D \times D$ matrix. The
structure of the matrices $\Gamma$ and $\Lambda$ is the same as in
the Matrix Product representation of pure states and the
TEBD-algorithm \cite{DMRG} can be employed for the simulation of
the dynamics. In contrast to pure states, the matrix elements of
the $\Lambda_j$ for mixed states can however no longer be
interpreted as the Schmidt coefficients of the respective
decomposition.

\section*{References}



\begin{thebibliography}{99}
\bibitem{Leggett}
A.J. Leggett and A. Garg, Phys. Rev. Lett. \textbf{54}, 857
(1985). \\
 A.J. Leggett, J. Phys.: Condens. Matter \textbf{14},
R415 (2002).
\bibitem{newleggett}
A.N. Jordan, A.N. Korotkov, M. B\"uttiker, Phys. Rev. Lett.
\textbf{97}, 026805 (2006).
\bibitem{reviews}
Y. Makhlin, G. Sch\"{o}n, A. Shnirman, Rev. Mod. Phys.
\textbf{73}, 357 (2001). \\
M.H. Devoret, A. Wallraff, J.M. Martinis, cond-mat/0411174.\\
J.~Q. You and F. Nori, Phys. Today \textbf{58}, 42 (2005).
\bibitem{Nakamura}
Y. Nakamura, Y.A. Pashkin, J.S. Tsai, Nature \textbf{398}, 786
(1999).
\bibitem{Mooij}
J.E. Mooij, T.P. Orlando, L. Levitov, L. Tian, C.H. van der Wal,
S. Lloyd, Science \textbf{285}, 1036 (1999).
\bibitem{Nori}
J.Q. You, J.S. Tsai, F. Nori, Phys. Rev. Lett. \textbf{89}, 197902
(2002).
\bibitem{Vion}
D. Vion, A. Aassime, A. Cottet, P. Joyez, H. Pothier, C. Urbina,
D. Esteve, M.H. Devoret, Science \textbf{296}, 886 (2002).
\bibitem{Yu}
Y. Yu, S. Han, X. Chu, S-I Chu, Z. Wang, Science \textbf{296}, 889
(2002).
\bibitem{Wallquist}
M. Wallquist, J. Lantz, V.S. Shumeiko, G. Wendin, New J. Phys.
\textbf{7}, 178 (2005).
\bibitem{Pashkin}
Y.A. Pashkin, T. Yamamoto, O. Astafiev, Y. Nakamura, D.V. Averin,
J.S. Tsai, Nature \textbf{421}, 823 (2003). \\
T. Yamamoto, Y.A. Pashkin, O. Astaviev, Y. Nakamura, J.S. Tsai,
Nature \textbf{425}, 941 (2003).
\bibitem{Majer}
J.B. Majer, F.G. Paauw, A.C.J. ter Haar, C.J.P.M. Harmans, J.E.
Mooij, Phys. Rev. Lett. \textbf{94}, 090501 (2005). \\
R. McDermott, R.W. Simmonds, M. Steffen, K.B. Cooper, K. Cicak,
K.D. Osborn, S. Oh, D.P. Pappas, J.M. Martinis , Science
\textbf{307}, 1299 (2005).
\bibitem{hybrid}
M. Paternostro, G.M. Palma, M.S. Kim, G. Falci, Phys. Rev. A
\textbf{71}, 042311 (2005). \\
P. Rabl, D. DeMille, J.M. Doyle, M.D. Lukin, R.J. Schoelkopf, P.
Zoller , Phys. Rev. Lett. \textbf{97}, 033003 (2006).
\bibitem{Romito}
A. Romito, R. Fazio, C. Bruder, Phys. Rev. B \textbf{71},
100501(R) (2005). \\
A. Lyakhov and C. Bruder, New J. Phys. \textbf{7}, 181 (2005).
\bibitem{Bose 03}
S. Bose, Phys. Rev. Lett. \textbf{91}, 207901 (2003).
\bibitem{discussion_Mooij}
J.E. Mooij, private communication (2005).
\bibitem{noise_circuit}
D. Loss and K. Mullen, Phys. Rev. A \textbf{43}, 2129 (1991). \\
P. Cedraschi and M. B\"uttiker, Phys. Rev. B \textbf{63}, 165312
(2001). \\
J.P. Pekola and J.J. Toppari, Phys. Rev. B \textbf{64}, 172509 (2001).\\
E. Almaas and D. Stroud, Phys. Rev. B \textbf{65}, 134502 (2002).\\
H. Kohler, F. Guinea, F. Sols, Ann. Phys. \textbf{31}, 127 (2004).
\bibitem{noise_spin-boson}
See A. Shnirman, Y. Makhlin, G. Sch\"{o}n, Phys. Scr. \textbf{T102},
147 (2002) and references therein.
\bibitem{noise_1/f}
Y. Nakamura,  Y.A. Pashkin, T. Yamamoto, J.S. Tsai, Phys. Rev.
Lett. \textbf{88}, 047901 (2002). \\
E. Paladino, L. Faoro, G. Falci, R. Fazio, \emph{ibid.}
\textbf{88}, 228304 (2002). \\
O. Astafiev, Y.A. Pashkin, Y. Nakamura, T. Yamamoto, J.S. Tsai,
\emph{ibid.} \textbf{93}, 267007 (2004). \\
L. Faoro, J. Bergli, B.L. Altshuler, Y.M. Galperin, \emph{ibid.}
\textbf{95}, 046805 (2005).\\
A. Shnirman, G. Sch\"{o}n, I. Martin and Y. Makhlin, \emph{ibid.}
\textbf{94}, 127002 (2005).\\
J. Schriefl, Y. Makhlin, A. Shnirman and G. Sch\"{o}n, New J. Phys.
{\bf 8}, 1 (2006).
\bibitem{noise_N}
M.J. Storcz and F.K. Wilhelm, Phys. Rev. A \textbf{67}, 042319
(2003). \\
J.Q. You,  X. Hu, F. Nori, Phys. Rev. B \textbf{72}, 144529
(2005).\\
B. Ischi, M. Hilke, M. Dub\'{e}, Phys. Rev. B \textbf{71}, 195325
(2005).
\bibitem{DMRG}
G. Vidal, Phys. Rev. Lett. {\bf 93}, 040502 (2004). \\
M. Zwolak and G. Vidal, Phys. Rev. Lett. \textbf{93}, 207205
(2004).
\bibitem{HRP06}
M.J. Hartmann, M.E. Reuter, M.B. Plenio, New. J. Phys. {\bf 8}, 94
(2006).
\bibitem{ent_review}
M.B. Plenio and S. Virmani, Quant. Inf. Comp. \textbf{7}, 1 (2007).
\bibitem{log_negativity}
J. Eisert, PhD thesis, University of Potsdam (2001). \\
M.B. Plenio, Phys. Rev. Lett. \textbf{95}, 090503 (2005).
\bibitem{APE03}
K. Audenaert, M.B. Plenio and J. Eisert, Phys. Rev. Lett.
\textbf{90}, 027901 (2003)
\bibitem{Martinis}
M. Steffen, M. Ansmann, R.C. Bialczak, N. Katz, E. Lucero, R.
McDermott, M. Neeley, E.M. Weig, A.N. Cleland, J.M. Martinis,
Science \textbf{313}, 1423 (2006).
\bibitem{AP06}
K. Audenaert and M.B. Plenio, New J. Phys. \textbf{8}, 266 (2006).
\bibitem{ent_generation}
J. Eisert,  M.B. Plenio, S. Bose, J. Hartley, Phys. Rev. Lett.
\textbf{93}, 190402 (2004).
\bibitem{Plenio HE 04}
M.B. Plenio, J. Hartley, J. Eisert, New J. Phys. \textbf{6}, 36
(2004)
\bibitem{discussion_Meeson}
P. Meeson, private communication (2006).
\bibitem{disorder}
S. Montangero and L. Viola, Phys. Rev. A \textbf{73}, 040302(R)
(2006). \\
S. Montangero, G. Benenti, R. Fazio, Phys. Rev. Lett. \textbf{91},
187901 (2003).
\bibitem{phonons}
L.B. Ioffe, V.B. Geshkenbein, Ch. Helm, G. Blatter, Phys. Rev.
Lett. \textbf{93}, 057001 (2004).
\bibitem{werner}
P. Werner, K. V\"{o}lker, M. Troyer, S. Chakravarty, Phys. Rev.
Lett. \textbf{94}, 047201 (2005).
\bibitem{hp}
S.F. Huelga and M.B. Plenio, quant-ph/0608164.
\bibitem{EBA06}
J. Eisert, F.G.S.L. Brand\~ao, K. Audenaert, quant-ph/0607167. \\
O. G\"{u}hne, M. Reimpell, R.F. Werner, quant-ph/0607163.
\bibitem{Bhatia}
R. Bhatia, Matrix Analysis, (Springer-Verlag, New York, 1997).

\end{thebibliography}
\end{document}